\documentclass[prl,twocolumn,superscriptaddress,showpacs,aps]{revtex4}

\usepackage{color}
\usepackage{graphicx}
\usepackage{bm}
\usepackage{amsmath}

\begin{document}

\newcommand{\Tm}{TmNi$_2$B$_2$C}
\newcommand{\Lu}{LuNi$_2$B$_2$C}
\newcommand{\Tc}{T_{\text{c}}}
\newcommand{\Hup}{H_{\text{c2}}}
\newcommand{\fq}{\phi_0}
\newcommand{\qo}{q_0}
\newcommand{\xiup}{\xi_{\text{c2}}}
\newcommand{\TN}{T_{\text{N}}}
\newcommand{\Jvec}{\textbf{J}}
\newcommand{\svec}{\textbf{s}}
\newcommand{\Hvec}{\textbf{H}}
\newcommand{\qvec}{\textbf{q}}
\newcommand{\rvec}{\textbf{r}}
\newcommand{\Avec}{\textbf{A}}
\newcommand{\kvec}{\textbf{k}}
\newcommand{\Bvec}{\textbf{B}}
\newcommand{\avec}{\textbf{a}}
\newcommand{\Mpara}{\textbf{M}_{\text{para}}}

\title{Pauli Paramagnetic Effects on Vortices in Superconducting \Tm}

\author{L. DeBeer-Schmitt}
\affiliation{Department of Physics, University of Notre Dame,
             Notre Dame, Indiana 46556, USA}

\author{M. R. Eskildsen}
\email{eskildsen@nd.edu}
\affiliation{Department of Physics, University of Notre Dame,
             Notre Dame, Indiana 46556, USA}

\author{M. Ichioka}
\affiliation{Department of Physics, Okayama University, Okayama 700-8530, Japan}

\author{K. Machida}
\affiliation{Department of Physics, Okayama University, Okayama 700-8530, Japan}

\author{N. Jenkins}
\affiliation{DPMC, University of Geneva, 24 Quai E.-Ansermet,
             CH-1211 Gen\`{e}ve 4, Switzerland}

\author{C. D. Dewhurst}
\affiliation{Institut Laue-Langevin, 6 Rue Jules Horowitz,
             F-38042 Grenoble, France}

\author{A. B. Abrahamsen}
\affiliation{Materials Research Department, Ris\o \ National Laboratory,
             DK-4000 Roskilde, Denmark}

\author{S. L. Bud'ko}
\affiliation{Ames Laboratory and Department of Physics and Astronomy,
             Iowa State University, Ames, Iowa 50011, USA}

\author{P. C. Canfield}
\affiliation{Ames Laboratory and Department of Physics and Astronomy,
             Iowa State University, Ames, Iowa 50011, USA}

\date{\today}

\begin{abstract}
The magnetic field distribution around the vortices in {\Tm} in the paramagnetic phase
was studied experimentally as well as theoretically.
The vortex form factor, measured by small-angle neutron scattering, is found to be
field independent up to $0.6 \Hup$ followed by a sharp decrease at higher fields.
The data are fitted well by solutions to the Eilenberger equations when paramagnetic
effects due to the exchange interaction with the localized 4$f$ Tm moments are
included.
The induced paramagnetic moments around the vortex cores act to maintain the field
contrast probed by the form factor.
\end{abstract}


\maketitle

The interplay between superconductivity and local magnetic moments is a fascinating
problem, with relevance to a number of important unresolved questions such as the
detailed nature of both high-$\Tc$ and heavy-fermion superconductivity.
Adding to this the usually antagonistic nature of superconductivity and magnetism, it
is no surprise that materials which exhibit a coexistence of these two goundstates
attract a lot of attention.

The antiferromagnetic members of the intermetallic nickelborocarbide superconductors
{\em R}Ni$_2$B$_2$C ({\em R} = Ho, Er or Tm) have proved especially rich vehicles for
such studies, displaying {\em e.g.} intertwined magnetic and superconducting
transitions as well as subtle changes in the superconducting characteristic length
scales associated with the onset of antiferromagnetic
ordering~\cite{Eskildsen98,Gammel99,Noergaard00}.
The exchange interaction ${\cal H}_{sf} = -I (g_J-1) \Jvec \cdot \svec$ between the
4$f$ localized moment $\Jvec$ and the conduction electron spin $\svec$ ($g_J$ is
Land\'{e} g-factor and $I$ is exchange integral) is important in understanding
systematic changes of both superconducting and magnetic transition
temperatures~\cite{Machida80}. However, even in the paramagnetic state above the
antiferromagnetic ordering temperature, $\TN$, the conduction electron moment
$\bm{\mu}_e = g \mu_B \svec$ is subjected to an exchange field
$\Hvec_{\text{ex}} = I (g_J-1) \Jvec /g\mu_B$ due to the field induced 4$f$-moments,
yielding a ``Zeeman'' term ${\cal H}_{sf} = - \Hvec_{\text{ex}} \cdot \bm{\mu}_e$ in
the conduction electron Hamiltonian.

Here we report on combined experimental and theoretical studies of \Tm, investigating
specifically how the magnetic field profile around the vortices is influenced by the
paramagnetic state.
Using small-angle neutron scattering (SANS) we imaged the vortex lattice (VL) at
several temperatures $> \TN$, and measured the magnetic field dependence of the
form factor which reflects the field distribution around the vortices. In contrast to
the usual exponential decrease with increasing field, the VL form factor in {\Tm}
remains constant up to $H \sim 0.6 \Hup$, followed by a sudden decrease as the upper
critical field is approached. It is the striking departure from exponential behavior
which is the central result of this Letter.

The experimental results are compared to solutions of the quasi-classical Eilenberger
equations, focusing on how the internal field distribution in the mixed state is
affected by changes to the electronic vortex core structure due to the paramagnetism.
Since $M(H)$ is roughly linear below $\Hup$ ~\cite{Cho95a} the induced moment $\Jvec$
is proportional to the applied field and can thus be treated as an effective Pauli
paramagnetic effect, giving rise to a Zeeman energy $\mu B$ where the parameter $\mu$
signifies the strength of the paramagnetic effect. The calculations show how the
induced moments in and around the vortices grow with increasing applied field and
thereby maintaining a high field modulation (and hence form factor), before they
eventually spread out from the core region at high fields. The results of the
calculations provide an excellent quantitative agreement with the measured form
factor.

{\Tm} has a superconducting critical temperature, $\Tc = 11$ K, and the Tm moments
order antiferromagnetically in a long-period transverse-modulated state below
$\TN = 1.5$ K~\cite{Eskildsen98,Cho95a,Lynn97,Sternlieb97}. At low temperature the
magnetic moments are along the $c$ axis, which consequently is the direction of the
maximum magnetic susceptibility~\cite{Cho95a,Lynn97}. For magnetic fields applied
parallel to the $c$ axis $\Hup$ shows a non-monotonic behavior, reaching a maximum
near $T = 5$ K and $\mu_0 \Hup = 1$ T due to the Tm sublattice magnetization,
decreasing upon approaching the magnetic ordering temperature and finally increasing
again below $\TN$~\cite{Cho95a,Eskildsen98}.
Previous studies showed simultaneous magnetic and VL symmetry transitions below $\TN$,
as well as peaks in the FLL reflectivity associated with the magnetic
transitions~\cite{Eskildsen98}.

SANS experiments were carried out at the D11 instrument at the Institut Laue-Langevin.
The {\Tm} single crystal used in the experiment was grown using a high temperature
flux method, using isotopically enriched $^{11}$B to reduce neutron
absorption~\cite{Canfield01}.
Incident neutrons with wavelengths of $\lambda_n$ = 6 - 8 {\AA} and a wavelength
spread of $\Delta\lambda_n/\lambda_n$ = 10$\%$ were used.
The VL diffraction pattern was collected by a position sensitive detector.
For all measurements, the sample was cooled in a horizontal magnetic field applied
parallel to the crystalline $c$ axis and the incoming neutrons.
Measurements obtained at zero field were used for background subtraction.

\begin{figure}[t]
  \includegraphics{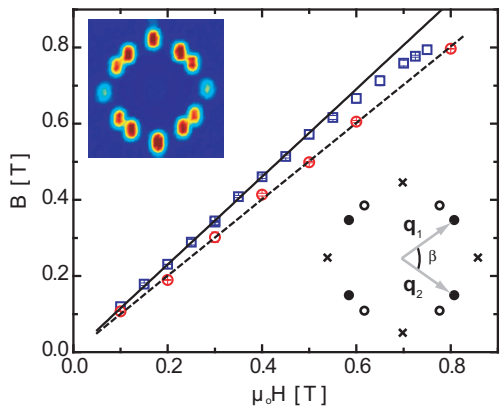}
  \caption{(Color online) Measured magnetic induction versus applied field at $1.6$ K
           for {\Tm} (squares) and non-magnetic LuNi$_2$B$_2$C (circles).
           The top left inset show a VL diffraction pattern obtained at $0.2$ T and
           $1.6$ K.
           The bottom right inset show a schematic of the diffraction pattern,
           indicating the VL scattering vectors and opening angle. Open and closed
           circles represent peaks belonging to different domain orientations, while
           $\times$'es denote higher order reflections.
           \label{BvsH}}
\end{figure}

The VL was imaged as a function of field at temperatures, $T = 1.6$ K, $3.5$ K and
$5.0$ K. At all fields and temperatures a rhombic VL was observed, as shown in the
insets to Fig.~\ref{BvsH}. The opening angle, $\beta$, was found to decrease with
increasing field indicating a continuous transition from a distorted square to a
distorted hexagonal symmetry in agreement with previous reports~\cite{Eskildsen98}. As
a consequence of having a non-square VL pinned to an underlying square crystalline
lattice, two VL domains were observed at all measured fields and temperatures.

A direct measure of the magnetic induction, $B$, in the sample can be obtained from
the VL scattering vectors. Using two scattering vectors belonging to the same domain,
the induction is given by
\begin{equation}
  B = \frac{\fq}{4\pi^2} | \qvec_1 \times \qvec_2 |,
  \label{induction}
\end{equation}
where $\fq = 20.7 \times 10^4$ T\AA$^2$ is the flux quantum.
Fig.~\ref{BvsH} shows the measured induction as a function of applied field for {\Tm}.
To rule out the possibility of systematic errors on the determination of $B$,
measurements on non-magnetic {\Lu} were performed immediately prior to the
measurements on {\Tm} using the same instrumental configuration. The measurements on
{\Lu} yielded $dB/d(\mu_0 H) = 1.003 \pm 0.006$ (dashed line) as expected for a
non-magnetic superconductor when $H \gg H_{\text{c1}}$.
For {\Tm} we find $B > \mu_0 H$ for the entire measured field range as seen in
Fig.~\ref{BvsH}, indicating a significant paramagnetic contribution to the induction.
Below $\sim 0.6$ T, $dB/d(\mu_0 H) = 1.152 \pm 0.004$ as indicated by the straight
line in Fig.~\ref{BvsH}. Taking demagnetization effects into account, this is in
excellent agreement with magnetization measurements~\cite{Cho95a}. At higher fields
$B$ approaches $\mu_0 H$, with the two fields seemingly merging at
$\mu_0 \Hup \approx 0.75$ T. We do presently not have an explanation for the high
field behavior of $B$.
\begin{figure}[t]
  \includegraphics{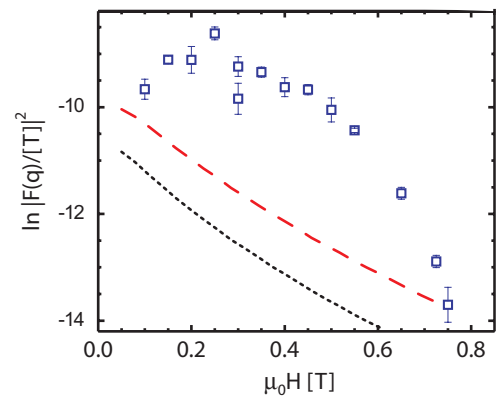}
  \caption{(Color online)
           Field dependence of the measured VL form factor in {\Tm} at $1.6$ K.
           The lines show the form factor calculated using eqn.~(\ref{Clem}) with
           $\xi = 210$ \AA \ and $\lambda = 780$ \AA \ (dotted line) or $600$ \AA \
           (dashed line). 
           \label{FF1}}
\end{figure}

We now turn to the main focus of this Letter: Measurements of the {\Tm} VL form
factor, $F(q)$, which is the Fourier transform of the magnetic field modulation due to
the vortices. Experimentally the form factor is related to the VL reflectivity by
\begin{equation}
  R = \frac{2 \pi \gamma^{2} \lambda_n^2 t}{16 \fq^2 q} |F(q)|^2,
  \label{reflec}
\end{equation}
where $\gamma = 1.91$ is the neutron gyromagnetic ratio, $t$ is the sample thickness,
and $q$ is the magnitude of the scattering vector~\cite{Christen77}.
Fig.~\ref{FF1} shows the VL form factor for {\Tm} at $1.6$ K just above $\TN$,
obtained from the integrated intensity of the Bragg peaks, as the sample is rotated
through the diffraction condition.

\begin{figure}[tb]
  \includegraphics{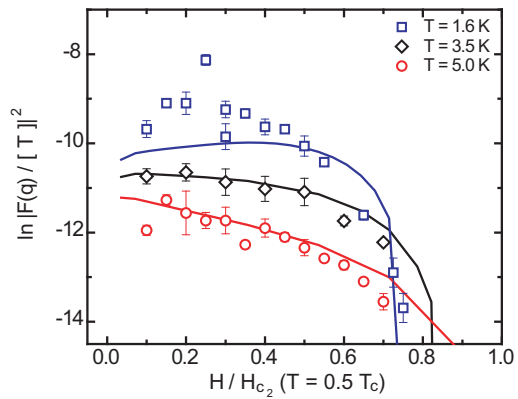}
  \caption{(Color online)
           Comparison of measured and calculated VL form factors in {\Tm} at
           $T = 1.6$, $3.5$, and $5.5$ K. The curves were calculated using the model
           described in the text, for $T = 0.16\Tc$ and $\mu = 1.71$ (A),
           $T = 0.35\Tc$ and $\mu = 1.28$ (B), and $T = 0.50\Tc$ and $\mu = 0.86$ (C).
           \label{FF2}}
\end{figure}

Using the model obtained by Clem and valid for
$\kappa_{\text{GL}} = \lambda/\xi \gg 1$~\cite{Clem75}:
\begin{equation}
  F(q) = B \frac{g \; K_{1}(g)}{1 + \lambda^2 q^2}, \hspace {1cm}
  g = \sqrt{2} \xi \; (q^2 + \lambda^{-2})^{1/2},
  \label{Clem}
\end{equation}
where $K_1$ is a modified Bessel function, we have calculated expected vortex form
factors shown by in Fig.~\ref{FF1}. The dotted line corresponds to a penetration
depth, $\lambda = 780$ \AA \ from literature~\cite{Cho95a} and a coherence length
based on the upper critical field at $1.6$ K,
$\xi_{c2} = \sqrt{\fq/2\pi \Hup} = 210$ \AA. As it is evident, the calculated form
factor falls substantially below our inferred values. Using a somewhat smaller value
of the penetration depth, $\lambda = 600$ \AA (dashed line), yields a better fit to
the endpoints of the measured form factor, but fails to describe the non-exponential
field dependence at all intermediate fields. Qualitatively similar results were also
obtained at temperatures further above $\TN$ as shown in Fig.~\ref{FF2}.
As we will show below the unusual field dependence of the form factor can be explained
by a microscopic calculation taking into account paramagnetic effects which modify the
magnetic field profile around the vortices. 

The form factor $F(\qvec_{h,k})$ is calculated from the internal field distribution 
$B(\rvec) = \sum_{h,k} F(\qvec_{h,k}) \exp(\text{i} \qvec_{h,k} \cdot \rvec)$
with the wave vector $\qvec_{h,k} = h \qvec_1 + k \qvec_2$,
$\qvec_1 = (2 \pi/a, -\pi/a_y,0)$ and $\qvec_2 = (2 \pi/a, \pi/a_y,0)$, 
corresponding to VL unit vectors $(a/2, a_y)$ and $(a/2, -a_y)$.  
The intensity of the main peak at $(h,k) = (1,0)$ gives the fundamental component
$F(\qvec_{0,1})$. To determine $B(\rvec)$ we selfconsistently calculate the spatial
structure of the pair potential $\Delta(\rvec)$ and the vector potential
$\Avec(\rvec)$ using the quasiclassical Eilenberger theory in the clean
limit~\cite{Eilenberger68,Klein87,Ichioka04,Klein00,Watanabe05,Ichioka07}, including
the paramagnetic contribution due to the effective Zeeman effect through the exchange
coupling of the conduction electron and {\Tm} sublattice moments. The quasi-classical
Green's functions $g( \omega_n, \kvec, \rvec)$, $f( \omega_n, \kvec, \rvec)$, and  
$f^\dagger( \omega_n, \kvec, \rvec)$ are calculated in the vortex lattice state by the
Eilenberger equations
\begin{eqnarray}
\left\{ \omega_n + \text{i} \mu B
   + \kvec \cdot \left( \nabla + \text{i} \Avec \right) \right\} f & = &
\Delta \phi(\kvec) g, 
\nonumber \\
\left\{ \omega_n + \text{i} \mu B  
   - \kvec \cdot \left( \nabla - \text{i} \Avec \right) \right\} f^\dagger & = &
\Delta^\ast \phi^\ast(\kvec) g,
\label{eq:Eil} 
\end{eqnarray}  
with $g = (1 - ff^\dagger)^{1/2}$, Re$\{g\} > 0$, the pairing function $\phi(\kvec)$,
Matsubara frequency $\omega_n = (2n+1)\pi T$, and effective Zeeman energy
$\mu B$~\cite{Ichioka07}.
Here $\mu$ determines the strength of the paramagnetic effect. A simple
two-dimensional Fermi surface is used, with a Fermi momentum unit vector given by
$\kvec = (\cos \theta, \sin \theta)$ and $0 \le \theta < 2\pi$.
With the magnetic field applied along the $z$ axis direction, the vector potential
$\Avec(\rvec) = \frac{1}{2} \bar{\Bvec} \times \rvec + \avec(\rvec)$ in the symmetric
gauge, where $\bar{\Bvec} = (0,0,\bar{B})$ is the average, uniform flux density and
$\avec(\rvec)$ is related to the modulated internal field such that
$\Bvec(\rvec) = \bar{\Bvec} + \nabla \times \avec(\rvec)$. 

The selfconsistent conditions for $\Delta(\rvec)$ and $\Avec(\rvec)$ are given by
respectively
\begin{equation}
\Delta =
   g_0 \, N_0 \, T \sum_{0 < \omega_l \le \omega_{\text{cut}}}
   \left\langle \phi^\ast(\kvec) \left(  
                f + {f^\dagger}^\ast \right) \right\rangle_{\kvec},  
\label{eq:scD}
\end{equation}
and
\begin{equation}
\nabla \times \left( \nabla \times \Avec \right) =
   \nabla \times \Mpara - \frac{2T}{\kappa^2}
   \sum_{0 < \omega_l} \left\langle \kvec \, \text{Im} \{g\}
                       \right\rangle_{\kvec},
\label{eq:scH}
\end{equation}
where $\langle \cdots \rangle_{\kvec}$ indicates the Fermi surface average, and
$\kappa = \sqrt{7 \zeta(3)/8} \kappa_{\text{GL}} \sim \kappa_{\text{GL}}$ where
$\kappa_{\text{GL}}$ is the Ginzburg-Landau (GL) parameter~\cite{Miranovic03}.
In Eq.~(\ref{eq:scD}), 
$(g_0 \, N_0)^{-1}=  \ln T +
                     2T \sum_{0 < \omega_l \le \omega_{\text{cut}}}\omega_l^{-1}$,
and we use $\omega_{\text{cut}} = 20 k_{\text{B}} \, \Tc$. In Eq.~(\ref{eq:scH}) both
the diamagnetic contribution of supercurrent in the last term and the contribution of
the paramagnetic moment
$\Mpara = (0,0,M_{\text{para}}(\rvec))$ with  
\begin{equation} 
M_{\text{para}}(\rvec) =
   \left( \frac{\mu}{\kappa} \right)^2
   \left( B(\rvec) - \frac{2T}{\mu} \sum_{0 < \omega_l}
          \left\langle \text{Im} \left\{ g \right\} \right\rangle_{\kvec} 
\right),
\label{eq:scM}  
\end{equation}
are treated fully self-consistently~\cite{Ichioka07}. As mentioned earlier, the
vortices in {\Tm} at low temperature and intermediate fields form a distorted square
VL~\cite{Eskildsen98}, indicating a large fourfold anisotropy of the Fermi surface and
pairing function~\cite{Nakai02}. We consequently use a pairing function
$\phi(\kvec) = | \sqrt{2} \cos2 \theta |$ and a square VL configuration ($a_y = a/2$).
However, the overall qualitative features of the form factor do not depend much on
these choices.

The field dependence of the calculated $|F(\qvec = \qvec_{0,1})|$ is shown in
Fig.~\ref{FF2}, where we have used $\mu_0 \Hup(T = 0.5\Tc) = 1$ T for comparison
with the experimental data.
\begin{figure*}[tb]
  \includegraphics{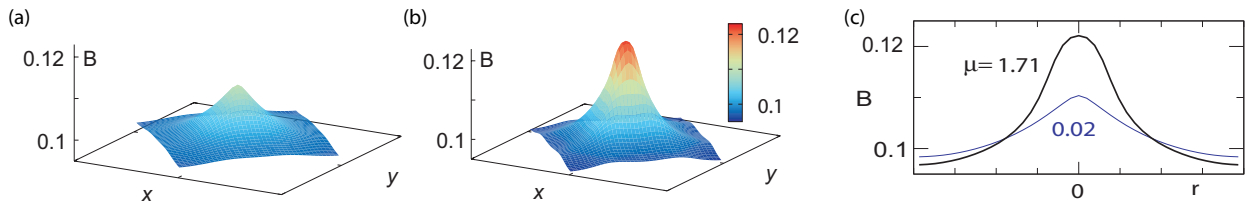}
  \caption{(Color online)
           Spatial structure of internal field $B(\rvec)$ within a unit cell of the
           square vortex lattice. 
           Here $\bar{B} = 0.1 = 0.36 \Hup(0.5\Tc)$, $\kappa = 6.2$, $T = 0.16\Tc$,
           and $\mu = 0.02$ (a) and $1.71$ (b). A profile of the field distribution is
           shown in (c).
           \label{Bprofile}}
\end{figure*}
The magnitude of $|F(q)|^2$ depends on the GL parameter, and its gradient as a
function of field is related to the paramagnetic parameter $\mu$. At $T = 0.5\Tc$
(5 K) values for $\kappa = 6.2$ and $\mu = 0.86$ were chosen to obtain agreement
between the calculated and measured form factor at low and intermediate fields.
The small deviation close to $\Hup$ may be due to the increasing deformation of the VL
away from a square symmetry ~\cite{Eskildsen98}, which is not included in this
calculation.
While $\kappa$ is kept constant for the remainder of the calculations, the value of
$\mu$ is expected to be proportional to the magnetization which in {\Tm} is dominated
by the contribution from the Tm 4$f$-moments~\cite{Cho95a}.
The decrease of $\Hup$ below 5 K can thus be attributed to an increasing paramagnetic
depairing. We therefore determine values of $\mu$ such that they reproduce the
suppression of $\Hup(T)$, yielding $\mu = 1.28$ at $T = 0.35\Tc$ ($3.5$ K) and
$\mu = 1.71$ at $T = 0.16\Tc$ ($1.6$ K), corresponding to respectively
$\Hup(T)/\Hup(0.5\Tc) \sim 0.85$ and $0.75$.
As a consequence of the increasing value of $\mu$, the slope of the form factor at
low fields changes from negative to positive (curves $C \rightarrow B \rightarrow A$
in Fig.~\ref{FF2}).
As evident from Fig.~\ref{FF2} the calculated form factor at $0.35\Tc$ provides a good
fit to the experimental data. At $0.16\Tc$ the calculated form factor captures the
qualitative field dependence, but falls below the datapoints at low fields. The reason
for this quantitative deviation is not clear, but it may be related to critical
behavior due to the close proximity to $\TN$. 

The paramagnetic moments induced around the vortex cores enhance $B(\rvec)$, and
consequently also the form factor $|F(\qvec_{01})|$ which acquires a paramagnetic
component proportional to $H$. To visualize the contribution from the paramagnetic
moments around the vortex cores, Fig.~\ref{Bprofile} shows the spatial structure of
the internal field $B(\rvec)$ in a VL unit cell for $\mu = 0.02$ (a) and $1.71$ (b).
The vortex field profile along the nearest neighbor direction is plotted in
Fig.~\ref{Bprofile}(c). This shows how the paramagnetic component is confined at the
vortex center resulting in the enhancement of the internal field. 

Before concluding we would like to emphasize the relative ``simplicity'' of {\Tm} as
well as the theoretical model used here to describe the SANS results. In contrast to
the $d$- or (triplet) $p$-wave pairing observed in respectively the high $\Tc$'s and
SrRuO$_4$~\cite{Maeno01} the Cooper pairs in {\Tm} are singlets with only a modest gap
anisotropy~\cite{Suderow01}. Likewise Pauli paramagnetic limiting and a possible
non-uniform superconducting (FFLO) state which have recently received considerable
attention in the heavy fermion superconductor CeCoIn$_5$~\cite{Matsuda07} is not
relevant in the case of {\Tm}~\cite{PL}. Instead we argue that one can consider {\Tm}
as a ``standard'' paramagnetic (above $\TN$) superconductor, thus providing a very
valuable reference for more exotic materials.

In summary, we have presented combined experimental and theoretical studies of
vortices and the vortex lattice in {\Tm} in the paramagnetic phase above $\TN$. 
The physical picture which emerges is that the conduction electron paramagnetic
moments induced by the exchange interaction accumulate exclusively around the vortex
cores, creating nano-tubes of Tm magnetization and maintaining the field distribution
contrast of the VL.
While our calculation used a simple model to describe the $H$- and $T$-dependences,
it was still able to capture the qualitative and quantitative behavior of the form
factor, emphasizing that paramagnetic effects are important in understaning the vortex
state in {\Tm}.

After submission we became aware of a theoretical paper by J. Jensen
and P. Hedeg\aa rd (to be published in Phys. Rev. B) which also treats the anomalous
field dependece of the form factor in \Tm.

\begin{acknowledgments}
We are grateful to V. G. Kogan for numerous stimulating discussions.
MRE acknowledges support by the Alfred P. Sloan Foundation.
Work at Ames Laboratory is supported by the U. S. Department of Energy, Basic Energy
Sciences under Contract No. W-7405-Eng-82.
\end{acknowledgments}


\newpage

\printfigures

\end{document}